# Enhanced Target Interaction Area of Helical Plasma Plumes in a Pulsed RF Atmospheric Plasma Jet

RADHIKA T. P.[1] and SATYANANDA KAR[1,*]

[1] Department of Energy Science and Engineering, Indian Institute of Technology Delhi, Hauz Khas, New Delhi 110016
[*]Corresponding author. satyananda@dese.iitd.ac.in

**Abstract.** Helical plasma plumes generated in pulsed radiofrequency (RF) atmospheric pressure plasma jets (APPJs) exhibit unique flow dynamics arising from the coupling between Kelvin-Helmholtz instabilities and baroclinic torque. While the fundamental mechanism responsible for helical plume formation has been established, the implications of this morphology for plasma-target interaction remain largely unexplored. In this study, we experimentally investigate the interaction area of helical plasma plumes with different target configurations and compare it with the conventional conical plume produced under continuous RF excitation. High-speed imaging reveals that the helical plume significantly enlarges the effective plasma-target interaction region due to its rotating trajectory and enhanced air entrainment. The effect of different target materials and boundary conditions, including dielectric surfaces and metal-backed substrates, is systematically examined. The results demonstrate that helical plumes provide superior surface coverage, improved spatial distribution of reactive species, and enhanced plasma-surface coupling. These findings highlight the advantages of helical plasma jets for applications requiring large-area plasma treatment, including plasma medicine, surface modification, and plasma-liquid interactions.

## 1. Introduction

Atmospheric pressure plasma jets (APPJs) have emerged as versatile tools for applications ranging from plasma medicine and surface modification to environmental remediation and nanomaterial processing [1, 2, 3, 4]. Their ability to generate reactive oxygen and nitrogen species (RONS) at near-ambient temperatures makes them particularly attractive for treatments involving heat sensitive biological and polymeric materials. Most APPJs produce a conical plasma plume, where the ionization front propagates along the laminar core of a noble gas jet expanding into ambient air [5, 6, 7, 8]. In such plumes, the interaction with a target surface is largely confined to a localized region directly beneath the jet axis. While this configuration provides stable plasma generation, the effective plasma-target interaction area is often limited, which can reduce treatment efficiency when uniform surface exposure is required.

Recently, a new class of plasma plume morphology has been observed in radiofrequency plasma jets operated under pulsed excitation. The earlier study demonstrated that pulsed RF operation can produce self-organized helical plasma plumes without the need for external magnetic fields or rotating electric fields [9]. Through Schlieren imaging and theoretical analysis, the formation of these helical structures was attributed to the combined effect of Kelvin-Helmholtz instability and baroclinic torque, which generate swirling gas flow structures that guide the plasma ionization front into a helical trajectory [10]. The helical plume geometry introduces several important changes to the plasma dynamics. First, the swirling motion increases entrainment of ambient air into the noble gas core, thereby enhancing the production of reactive species. Second, the plasma column no longer propagates along a single axial direction but instead traces a rotating path in space. This motion effectively sweeps a larger spatial region compared to the stationary conical plume. Despite these promising characteristics, the implications of helical plume morphology for plasma-target interactions have not yet been systematically explored. In many practical applications, such as plasma medicine, surface treatment, and plasma-liquid processing, the effectiveness of plasma treatment depends strongly on the area and uniformity of plasma exposure.

The helical plasma plume investigated in the present study originates from fluid-dynamic instabilities within the underlying argon jet. In our previous work, Schlieren imaging and theoretical analysis demonstrated that velocity shear at the argon-air interface generates Kelvin-Helmholtz vortices under both continuous and pulsed RF excitation.

Under continuous operation, these vortices develop downstream of the luminous plasma region, resulting in the conventional conical plume. Under pulsed RF excitation, periodic thermal forcing generates strong pressure oscillations and baroclinic torque near the nozzle exit, causing the vortices to acquire azimuthal motion and produce a swirl-structured gas flow. The plasma ionization front follows this swirling trajectory, leading to the formation of a self-organized helical plume. A detailed analysis of the instability mechanism has been reported elsewhere [10]. The present work focuses on how this unique plume morphology influences plasma-target interaction characteristics.

The objective of the present work is therefore to investigate how the helical plume structure influences plasma-target interaction area under different surface conditions. The selected target configurations were chosen to represent a broad range of practical plasma-processing environments. Dielectric surfaces are representative of polymers, dielectric coatings, biological tissues, and other electrically insulating materials commonly encountered in plasma medicine and surface functionalization. Metallic targets are relevant to plasma-assisted cleaning, activation, coating, and treatment of conductive materials. Plasma-liquid interactions are of particular interest for plasma medicine, plasma-activated water generation, environmental remediation, and agricultural applications. In addition, dielectric-covered metallic substrates represent technologically important multilayer systems encountered in biomedical devices, coated materials, dielectric barrier discharge configurations, and flexible electronic platforms. Investigating these different target conditions therefore provides insight into how plume morphology influences plasma interaction across a wide range of practical applications. By comparing helical and conical plumes interacting with various target configurations, including dielectric and metal-backed surfaces, we provide experimental evidence demonstrating the advantages of helical plasma jets for surface treatment applications.

## 2. Experimental Set up

A schematic representation of the experimental setup used for plasma generation, and plasma-target interaction studies is shown in Fig. 1. The plasma jet system used in this study is identical to that described in our previous work on helical plume formation [10, 11]. The discharge source consisted of a quartz tube with an inner diameter of 6 mm and a wall thickness of 2 mm, which served as the dielectric channel for the plasma jet. A copper rod electrode of 1.6 mm diameter was inserted along the central axis of the tube and connected to the RF power supply, acting as the powered electrode. A copper strip of approximately 3 mm width was wrapped externally around the quartz tube near the exit region to serve as the grounded electrode, thereby forming a rod-ring electrode configuration. This geometry allows plasma ignition within the tube while enabling the plasma plume to extend downstream into the surrounding atmosphere. The discharge was driven by a 13.56 MHz radio frequency (RF) generator coupled through an impedance matching network to maintain a matched load of 50 Ω. Argon gas was used as the working gas and was introduced through the quartz tube at controlled flow rates using a mass flow controller. Under RF excitation, plasma was initiated between the central powered electrode and the external grounded electrode, producing a stable atmospheric pressure plasma jet that propagated beyond the tube exit into ambient air.

Two different excitation modes were employed in order to generate distinct plume morphologies. In the first case, the plasma jet was operated in continuous RF mode, which produced the conventional conical plasma plume commonly observed in atmospheric pressure plasma jets. In the second case, the RF power was modulated using a pulse generator to produce pulsed RF excitation with a frequency of 500 Hz and a duty cycle of 50%. Under these conditions, the plasma plume developed into a stable helical structure, as previously reported [9, 10]. The pulsed excitation induces periodic heating and cooling of the gas flow, which plays a key role in triggering the fluid dynamic instabilities responsible for the formation of the helical plume. The operating conditions used in the present study were selected based on our previous investigation of helical plume formation [10], where the influence of gas flow rate and RF power on plume morphology and stability was systematically examined. That study identified an operating window in which stable and reproducible helical plumes were generated. To isolate the effect of target electrical boundary conditions on plasma-surface interaction, all experiments reported here were performed under fixed

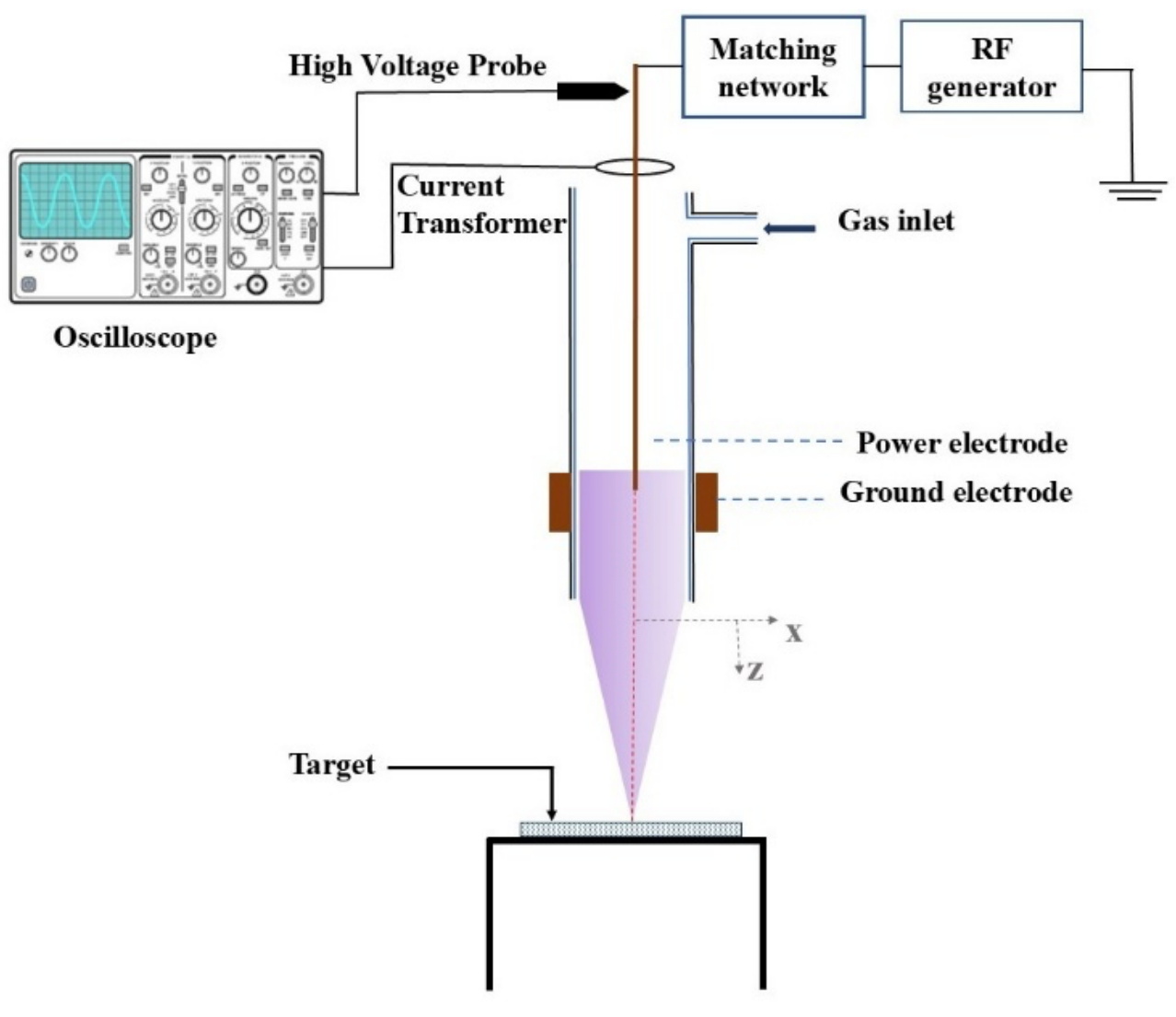

*Figure 1: Schematic of the experiment setup*

operating conditions within this stable helical-plume regime.

The interaction between the plasma plume and different target materials was investigated under identical discharge conditions in order to compare the treatment characteristics of the conical and helical plumes. The targets included dielectric substrates, metallic surfaces, and composite configurations consisting of dielectric layers placed on top of a metallic base. Such configurations were selected to examine how the electrical boundary conditions of the target influence plasma-surface interaction. The distance between the nozzle exit and the target surface was maintained constant throughout the experiments to ensure consistent interaction geometry. The spatial interaction between the plasma plume and the target surface was recorded using high-speed imaging, enabling visualization of the temporal evolution of the plasma footprint during plasma-surface interaction. The effective interaction area reported in this work corresponds to the cumulative plasma-surface contact region obtained over a fixed observation interval rather than the instantaneous plasma-channel diameter at a single time frame. Successive image frames were processed using ImageJ software to determine the total spatial region contacted by the plasma plume during the observation period. Prior to image analysis, all recorded high-speed images were spatially calibrated using the known outer diameter of the quartz tube as a reference dimension. The pixel-to-length conversion factor was determined using ImageJ software, enabling all plume dimensions and plasma-target interaction areas to be expressed in physical units (mm and $mm^2$). The effective interaction areas reported in Table 1 were obtained after applying this spatial calibration to the processed image sequences.

The electrical behavior of the discharge was characterized by monitoring the instantaneous voltage and current waveforms using a calibrated [12] high-voltage probe and current transformer connected to a digital storage oscilloscope. The absorbed power was determined directly from the phase-resolved voltage and current measurements in order to distinguish the actual plasma power dissipation from the nominal RF generator output. As discussed in [10], the generator input power differs significantly from the net delivered power due to impedance mismatch and reactive circuit components within the plasma jet assembly. Under continuous RF operation, generator powers of 55-105 W corresponded to absorbed powers of approximately 20.9-50.3 W, whereas under pulsed RF excitation (500 Hz, 50% duty cycle), the corresponding absorbed powers were substantially lower, approximately 8.9-24.3 W. For consistency, the comparative plasma-target interaction experiments presented in this work were performed under conditions corresponding to approximately equal average delivered power. These measurements confirm that the helical plume generated under pulsed RF excitation can achieve substantially enhanced interaction area and spatial plasma coverage at lower average dissipated power compared with the conventional continuous RF mode.

## 3. Results and Discussion

### *3.1 Plasma-Target Interaction Area*

The interaction between an atmospheric pressure plasma jet and a target surface is governed by the combined dynamics of the ionization wave, the neutral gas flow, and the electric field distribution near the surface [13, 14]. In the case of a conventional conical plasma plume produced under continuous RF excitation, the plasma column propagates predominantly along the central axis of the gas jet. The ionization front remains confined within the laminar core of the argon flow, which acts as a guiding channel for the plasma discharge [5]. As a result, when the plume interacts with a target surface, the effective plasma-target interaction region is primarily determined by the cross-sectional diameter of the plume at the point of impact. Since the plasma propagation follows the axial flow of the gas jet, the interaction footprint on the surface remains localized around the jet axis. Although some radial spreading occurs due to diffusion of reactive species and weak turbulent mixing with the surrounding air, the majority of the plasma activity remains concentrated within a relatively narrow region. Consequently, the treatment area produced by a conical plasma plume is typically limited, which can restrict its effectiveness in applications requiring large-area surface modification.

**Table 1. Area of contact for helical plume and conical plume for different target conditions**

| Target conditions | Spread over the target (mm) | | Area of contact ($mm^2$) | |
|---|---|---|---|---|
| | **Helical Plume** | **Conical Plume** | **Helical Plume** | **Conical Plume** |
| Dielectric | 7.084 | 5.907 | 39.393 | 27.390 |
| Floating metal | 4.772 | 2.804 | 17.876 | 6.171 |
| Floating metal with dielectric cover | 10.132 | 3.79 | 80.586 | 11.275 |
| Grounded metal | 4.131 | 3 | 13.396 1 | 7.065 |
| Grounded metal with dielectric cover | 10.426 | 5.016 | 85.330 | 19.750 |
| Liquid | 4.525 | 2.902 | 16.073 | 6.610 |
| Liquid with floating metal base | 5.164 | 3.64 | 20.933 | 10.400 |
| Liquid with grounded metal base | 7.082 | 3.295 | 39.371 | 8.522 |

A markedly different behaviour is observed when the plasma jet is operated under pulsed RF excitation, where the plume adopts a helical morphology. In this case, the plasma column no longer propagates along a single axial direction but instead follows a rotating spiral trajectory. The helical motion arises from the swirling gas flow generated near the nozzle exit, which continuously deflects the ionization front away from the central axis [10]. As the plasma column propagates downstream, it traces a helical path in space, effectively sweeping across a wider spatial region. When this rotating plasma plume interacts with a target surface, the plasma footprint becomes significantly larger compared with that produced by a conventional conical plume. Instead of interacting with the surface at a single point along the jet axis, the helical plume periodically contacts different regions of the surface as it rotates. This dynamic interaction results in a broader treatment footprint and a more uniform spatial distribution of plasma exposure. It should be noted that the enhancement reported here primarily reflects the cumulative interaction area integrated over time rather than a substantial increase in the instantaneous plasma-channel diameter. In both conical and helical plume configurations, the instantaneous plasma attachment near the surface remains relatively localized. However, while the conical plume remains approximately stationary near the jet axis, the helical plume continuously rotates and shifts its interaction position across the target surface. As a result, the total plasma-treated region accumulated over the observation interval becomes significantly larger for the helical plume configuration.

To quantify the influence of plume morphology on plasma-surface interaction, the effective interaction area of both conical and helical plumes was determined from high-speed imaging data. The plasma-target contact region was extracted from the recorded frames and analyzed using the ImageJ software. The resulting interaction areas for different target configurations are summarized in Table 1. The difference in plasma-target interaction patterns between conical and helical plumes is clearly visible in the images shown in Fig.2 and Fig. 3, where the helical plume produces a substantially larger cumulative interaction footprint over time due to its rotating trajectory across the target surface.

These results demonstrate that the helical plume consistently produces a larger interaction area than the conical plume across all investigated target configurations. The enhancement ranges from approximately 40% to 80%, depending on the electrical properties of the target surface. This systematic increase confirms that plume morphology plays a critical role in determining the efficiency of plasma-surface interaction.

### *3.2 Fluid Dynamic Origin of Helical-Plasma Surface Interaction*

The enhanced interaction area observed for the helical plume can be understood in terms of the fluid dynamics governing the plasma jet. As demonstrated in our previous study [10], pulsed RF excitation introduces periodic heating and cooling cycles within the plasma channel. These thermal oscillations significantly perturb the neutral gas flow, leading to the disruption of the laminar potential core that normally stabilizes the plasma plume in continuous operation [15]. Under pulsed excitation, the periodic forcing associated with RF modulation results in a high Strouhal number, which promotes the development of Kelvin-Helmholtz (K-H) instability directly at the nozzle exit [16]. The primary instability originates at the shear layer formed between the argon jet and the surrounding ambient air. Small perturbations at this interface are amplified by the velocity difference between the two gases, leading to the formation of Kelvin-Helmholtz vortices. It is important to note that K-H vortices are also present under continuous RF operation; however, in that case, the vortices develop farther downstream and remain largely outside the luminous plasma region, resulting in the conventional conical plume morphology. Under pulsed RF excitation, the periodic thermal forcing shifts the onset of instability toward the nozzle exit and significantly strengthens vortex growth within the plasma-active region.

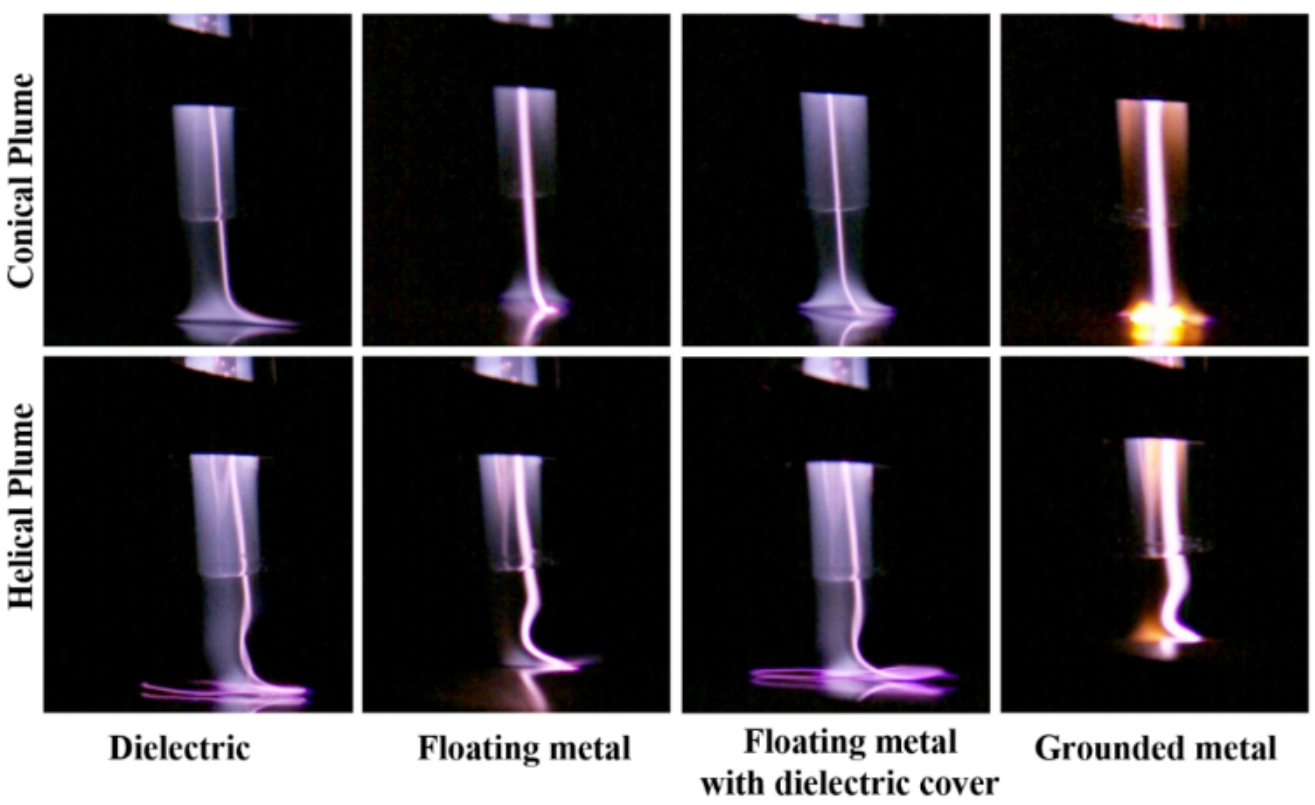

*Figure 2: High-speed images showing plasma-target interaction for conical and helical plumes under different target conditions. All images are spatially calibrated. The outer diameter of the quartz tube (8 mm) was used as the spatial calibration reference.*

For the operating conditions used in this study, the characteristic argon flow velocity is approximately 2.358 m $s^{-1}$, corresponding to a transit time of roughly 8-12 ms through the active plume region. Since the RF modulation frequency is 500 Hz (period = 2 ms), individual gas parcels experience multiple heating and cooling cycles during their propagation through the discharge. These repeated thermal perturbations generate transient temperature, density, and pressure gradients within the plasma-heated argon core. Although the instability originates at the argon-air interface, the resulting thermal gradients strongly influence its subsequent evolution.

In addition to the K-H instability, plasma heating generates strong temperature gradients within the jet, which produce corresponding gradients in gas density and pressure. When the density and pressure gradients become misaligned, a baroclinic torque is generated according to [17]:

$$\frac{d\boldsymbol{\omega}}{dt} \propto \nabla\boldsymbol{\rho} \times \nabla\boldsymbol{P}$$

where ω represents the vorticity of the flow. This baroclinic torque twists the K-H vortices and introduces an azimuthal component to the gas motion, thereby producing a swirling flow structure. Because the plasma ionization front is strongly coupled to the underlying neutral gas dynamics, the plasma column follows the swirling trajectory of the gas flow [10]. The result is the formation of a stable helical plasma plume.

It should be noted that buoyancy-driven convection is expected to play only a minor role under the present operating conditions. In our previous study [10], Richardson number analysis yielded values significantly below the critical threshold for buoyancy-dominated flow, indicating that the observed vortex formation is governed primarily by shear-driven instability and baroclinic vorticity generation rather than natural convection. Consequently, the dominant mechanism responsible for plume rotation is the combined action of Kelvin-Helmholtz instability and baroclinic torque acting on the plasma-heated argon jet. As this rotating plasma column propagates toward the target surface, it effectively scans across the surface area, thereby increasing the spatial coverage of plasma exposure and producing the enlarged plasma-target interaction footprint observed in the present experiments.

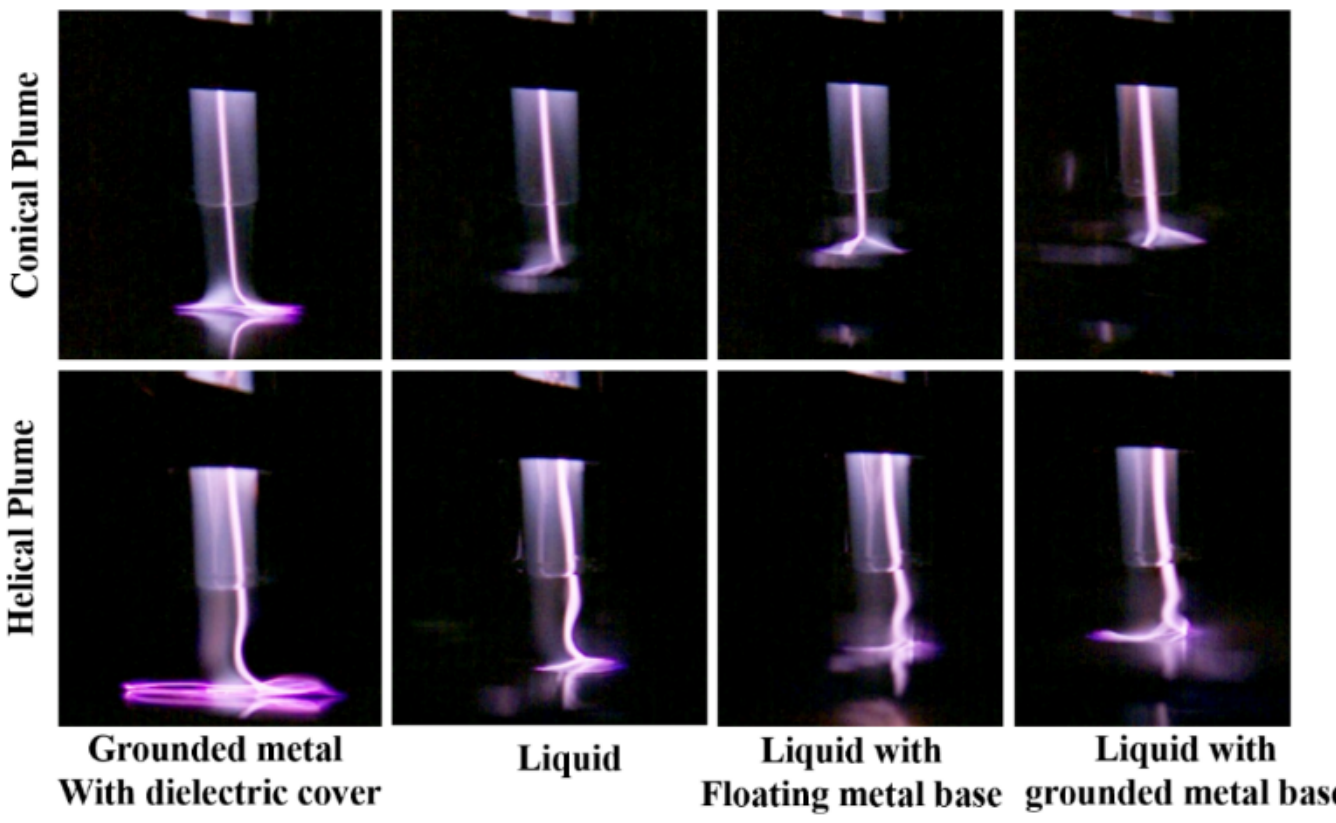

*Figure 3: High-speed images showing plasma-target interaction for conical and helical plumes under different target conditions. All images are spatially calibrated. The outer diameter of the quartz tube (8 mm) was used as the spatial calibration reference.*

### *3.3 Influence of Target Electrical Boundary Condition*

The nature of the target surface plays a crucial role in determining the characteristics of plasma-surface interaction. The electrical properties of the target material influence the distribution of electric fields near the surface, which in turn affects the propagation of the plasma ionization front and the spatial extent of the interaction region. When the plasma plume interacts with a dielectric surface, charge accumulation occurs due to the deposition of electrons and ions from the plasma. This charge buildup creates a localized surface potential that modifies the electric field distribution near the target. The resulting electric field redistribution can influence the trajectory of the ionization front and regulate the local plasma intensity. In many cases, the presence of surface charge tends to stabilize the plasma-surface interaction and promote a more uniform distribution of plasma exposure across the surface. In contrast, metallic targets provide a conductive boundary that allows surface charges to dissipate rapidly. The absence of charge accumulation prevents the formation of significant surface potentials, resulting in a stronger coupling between the plasma plume and the conductive surface. This enhanced coupling can increase the local electric field strength near the surface and intensify plasma activity at the interaction point. However, because the interaction remains localized along the jet axis for a conical plume, the treatment area remains relatively small.

From an electrical perspective, the target boundary condition also influences the current transport pathway between the plasma plume and the surface. A grounded metallic target provides a conductive path for charge transport and can therefore act as an effective charge sink for electrons and ions arriving from the plasma. This modifies the local electric field distribution and generally strengthens plasma-surface coupling. In contrast, dielectric surfaces accumulate deposited charges because no continuous conduction path exists through the material. The resulting surface charge generates an opposing electric field that progressively limits further charge deposition. Consequently, plasma interaction with dielectric targets is governed primarily by surface charging and displacement-current effects rather than direct conductive current flow. When a dielectric layer is placed on top of a metallic substrate, both mechanisms contribute simultaneously. The dielectric stores surface charge while the underlying conductive layer modifies the electric field through image-charge effects. This combination can enhance lateral plasma spreading and increase the effective interaction area, particularly for the rotating helical plume configuration.

An interesting intermediate case arises when a dielectric layer is placed on top of a metallic substrate. In such configurations, the dielectric layer acts as a charge-storage medium, while the underlying metal influences the electric field distribution through image charge effects. The presence of the conductive base effectively enhances the electric field within the dielectric layer, allowing charge accumulation to occur while simultaneously modifying the plasma-surface coupling. For helical plasma plumes, this configuration is particularly advantageous because the rotating plasma column interacts with different regions of the dielectric surface as it sweeps across the target. The experimental observations summarized in Table 1 confirm that, for all investigated target configurations, the helical plasma plume produces a larger interaction footprint than the conventional conical plume. A schematic summary of the influence of different target electrical boundary conditions on charge transport, electric field distribution, and plasma-surface interaction is presented in Fig. 4. Taken together, these results indicate that the combination of helical plume dynamics and favourable electrical boundary conditions can significantly increase the effective plasma-target interaction area. The largest interaction area measured in this study, 85.3 $mm^2$, is obtained for the helical plume interacting with a grounded metallic base with a dielectric cover. This value is more than 93% larger than the interaction area obtained for the conical plume interacting with a floating metal base without dielectric cover, highlighting the strong influence of both plume morphology and target electrical properties on plasma-surface interaction.

### 3.4 Implications for Plasma Processing Applications

The increased plasma-target interaction area produced by helical plumes offers several significant advantages for plasma-based surface processing applications. One of the primary benefits is the ability to achieve more uniform surface treatment without requiring mechanical scanning of the plasma source. Because the helical plume naturally sweeps across the surface as it propagates, it effectively distributes plasma exposure over a larger region. This characteristic is particularly advantageous for applications such as polymer activation, sterilization, biomedical therapy, and thin-film surface modification, where uniform plasma exposure is essential for consistent treatment results. Another important advantage arises from the enhanced mixing between the plasma effluent and the surrounding ambient air. The swirling motion of the helical plume increases the entrainment of air molecules into the plasma region, which promotes the formation of reactive oxygen and nitrogen species (RONS). These species play a central role in many plasma-assisted chemical processes, including surface functionalization and microbial inactivation. Furthermore, the helical plume geometry increases the effective plasma-air interaction surface, thereby extending the reaction pathway for plasma-generated species. This extended interaction pathway allows for more efficient chemical conversion processes and increases the overall reactivity of the plasma plume. Taken together, these characteristics demonstrate that helical plasma plumes provide a highly effective configuration for applications requiring large-area plasma treatment and enhanced plasma-surface interaction.

An additional advantage of the RF-modulated helical plume is its improved power utilization efficiency. Compared with continuous RF operation, pulsed excitation generates the helical plume at significantly lower average absorbed power while simultaneously increasing the effective plasma-target interaction area. This behavior arises because pulsed operation modifies the underlying plasma-fluid coupling rather than relying solely on increased power deposition to enlarge the treatment region. The rotating helical trajectory enables the plasma column to sweep across a larger spatial region without requiring mechanical scanning or increased discharge power. Consequently, RF modulation provides an energy-efficient approach for achieving large-area plasma treatment with enhanced spatial uniformity and reduced thermal loading. Such characteristics are particularly advantageous for temperature-sensitive applications including plasma medicine, polymer treatment, and plasma-liquid interaction systems.

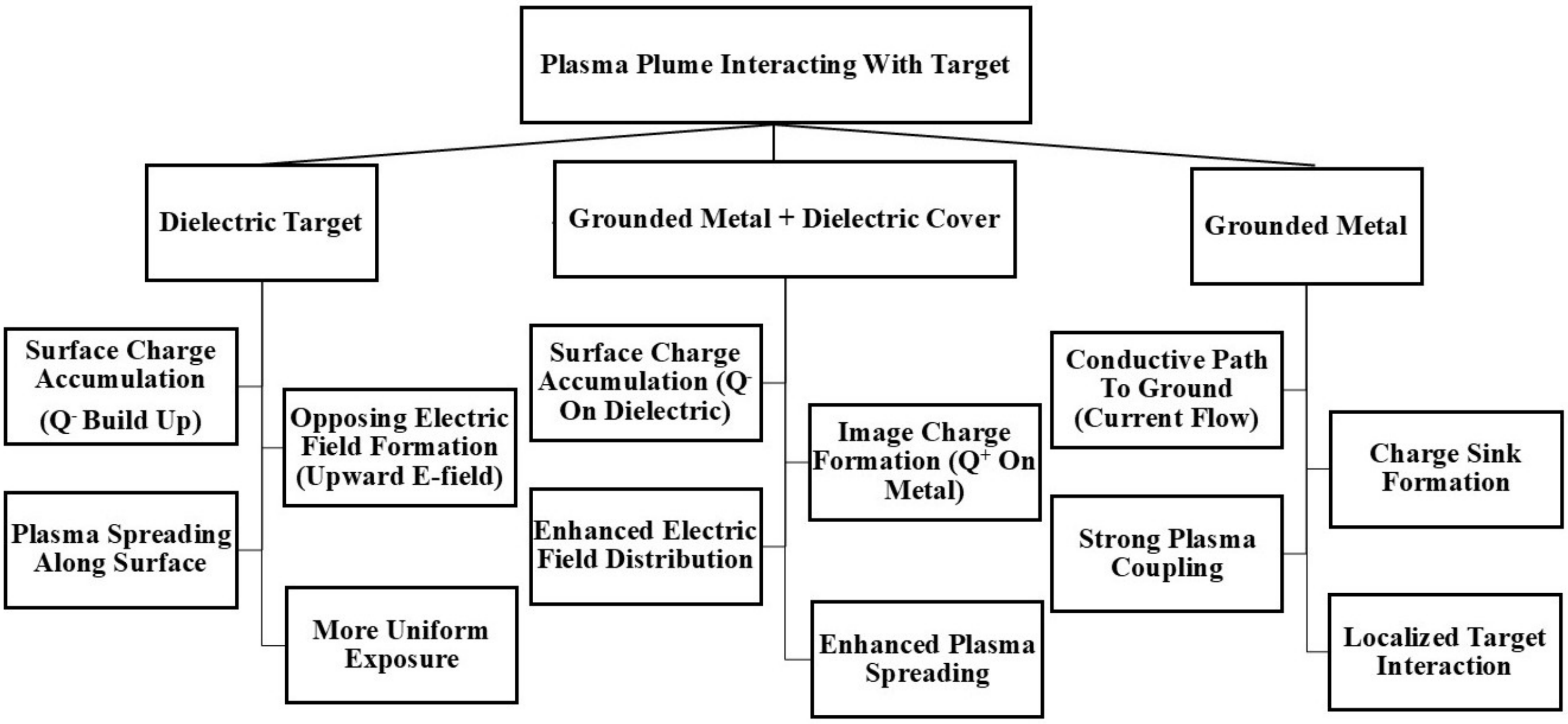

*Figure 4: Flow chart illustrating the influence of target electrical boundary conditions on plasma-surface interaction. Dielectric targets promote surface charging and plasma spreading, grounded metallic targets provide conductive current pathways and stronger plasma-surface coupling, while dielectric-covered metallic targets combine surface charging and image-charge effects, leading to enhanced electric-field distribution and larger interaction areas.*

The enhanced interaction area generated by the helical plume may be particularly beneficial in applications where treatment uniformity is critical. In plasma medicine, the enlarged treatment footprint could improve the coverage of biological tissues without requiring mechanical scanning of the plasma source. For polymer activation and surface functionalization, the rotating plume may provide more uniform modification over larger surface areas. In plasma agriculture and plasma-activated water production, the increased interaction region can enhance plasma-liquid contact and improve reactive species transfer into the liquid phase. Similarly, in sterilization, environmental remediation, and plasma-assisted materials processing, the self-scanning behavior of the helical plume offers the possibility of treating larger areas while maintaining relatively low average power consumption. These characteristics highlight the potential of helical plasma jets as a versatile platform for large-area plasma processing applications.

## 4. Conclusion

In this work, we investigated the influence of plasma plume morphology on plasma-target interaction characteristics in a radiofrequency atmospheric pressure plasma jet (RF-APPJ). By comparing the interaction of conventional conical plumes generated under continuous RF excitation with helical plumes formed under pulsed RF operation, we quantified how plume structure affects the effective treatment area on different target configurations. The interaction areas were extracted from high-speed imaging data and analyzed using ImageJ software, enabling a direct comparison of the plasma-surface footprint for various target conditions.

The experimental results demonstrate that the helical plasma plume consistently produces a larger interaction area than the conventional conical plume. The physical origin of this enhancement can be understood in terms of the coupled plasma-fluid dynamics governing the jet morphology. Under pulsed RF excitation, periodic heating and cooling cycles disrupt the laminar core of the argon jet, leading to the onset of Kelvin-Helmholtz instability near the nozzle exit. The resulting vortical structures, combined with baroclinic torque arising from misaligned pressure and density gradients, generate a swirling neutral gas flow. Because the ionization front is strongly coupled to the neutral gas dynamics, the plasma column follows this swirling trajectory and forms a helical plume. As the plasma column propagates toward the target surface, the rotating ionization front continuously shifts its point of contact with the surface, effectively scanning across a larger spatial region. This dynamic interaction leads to the significant enlargement of the plasma-target interaction footprint observed in the experiments.

In addition to plume dynamics, the electrical properties of the target surface also play an important role in determining the interaction area. Dielectric targets promote charge accumulation, which modifies the local electric field distribution and allows the plasma to spread laterally along the surface. When a dielectric layer is supported by a metallic substrate, image charge effects further enhance the electric field near the surface, strengthening the plasma-surface coupling. The combination of these effects results in the largest interaction area observed in this study for the dielectric–metal configuration. The helical plume benefits particularly strongly from this configuration because its rotating trajectory allows the plasma column to interact sequentially with different regions of the surface.

These findings demonstrate that plume morphology is a key parameter in controlling plasma–surface interaction in atmospheric pressure plasma jets. Compared with conventional conical plumes, helical plumes provide larger treatment areas, improved spatial coverage, and enhanced plasma-surface coupling without requiring mechanical scanning of the plasma source. The increased interaction area is expected to promote more efficient transport of reactive species and improve the uniformity of plasma treatment across the target surface. Overall, the present results highlight the potential of helical plasma plumes as an effective configuration for applications requiring large-area plasma exposure, including plasma medicine, surface functionalization, sterilization, and plasma-assisted materials processing. Future work will focus on quantitative measurements of reactive species flux and detailed modelling of plasma-surface interaction under helical plume conditions in order to further optimize this configuration for practical plasma applications. Although the present work focuses on a single operating condition corresponding to a stable helical plume regime, the interaction area is expected to depend on gas flow rate and RF power through their influence on plume diameter, vortex strength, and plasma-fluid coupling. A systematic parametric investigation of these effects will be the subject of future studies aimed at optimizing helical plasma jets for specific application requirements.

## Acknowledgement

Radhika T.P. sincerely acknowledges the insightful contributions and support from Aishik Basu Mallick, Tejashwi Rana, Suryasunil Rath, and Pratyay Chattopadhyay during the course of this research.